\documentclass[a4paper]{article}
\usepackage{graphicx}

\begin{document}

\begin{center}
{\bf \Large
The norm game in a mean-field society}
\bigskip

{\large
K. Ku{\l}akowski$^{\dag}$
}
\bigskip

{\em
Faculty of Physics and Applied Computer Science,
AGH University of Science and Technology,
al. Mickiewicza 30, PL-30059 Krak\'ow, Poland\\

}

\bigskip

$^\dag${\tt kulakowski@novell.ftj.agh.edu.pl}

\bigskip
\today
\end{center}

\begin{abstract}
Mean field Master equations for the so-called norm game are proposed. The strategies are: 
to obey the norm or not and to punish those who break it or not. The punishment, the temptation,
the punishment cost and the relaxation of vengeance are modeled by four parameters;
for the fixed points, only two ratios of these parameters are relevant. The analysis
reveals two phases; in one of them, nobody obeys the norm and nobody punishes. 
This phase is stable if the punishment is small enough. In the other phase, the proportion
of defectors depends on the parameters and in some cases it can be arbitrarily small.
A transcritical bifurcation appears between the two phases. Numerical calculations show 
that the relaxation time shows a sharp maximum at the bifurcation point. The model is adapted 
also for the case of two mutually punishing groups. A difference between the solutions 
for two groups appears if the punishment of one group by the other is weaker, than the opposite.
\end{abstract}



\bigskip

\section{Introduction}

The game theory went to the statistical physics \cite{szafat} as a strong mathematical tool of economy \cite{neumorg,tsch}, biology \cite{hofsig} 
and social sciences \cite{ax2}. Initially, it relied on the assumption of rational choice. This condition was later released by the 
concept of bounded rationality \cite{simbo} and adaptive rather than rational thinking; this should not be understood as 'less logic', but
rather 'logic with limited information'. In the never-ending discussion on the applicability of mathematical tools to describe the human behavior,
this adaptive thinking is a keyword. Indeed, in many cases the human behavior cannot be explained within the frames of individual rationality.
If a society persists longer than the lifetime of its members, it develops ways to induce cooperation and altruism in further generations; 
examples of individual sacrifices constitute tradition and serve as social norms. As it was formulated by Elias J. Bickerman, "The first need of any social system is to create incentives to make people do more work than that required by their immediate wants" \cite{colu}.

The norm game was introduced by Axelrod \cite{axeng} together with a metanorm game, where punishing for non-punishing was included.
Axelrod presented some results of the computer simulations, which were later questioned \cite{gaizq}. On the other hand, recently reported 
results on the norm game \cite{hau} relied on a definite sets of values of the model parameters. It is then difficult to evaluate to what
extent these results are generic. This obstacle is even more painful when we realize that it concerns also the dependence of the results on 
a particular model, which is many cases is chosen arbitrarily. For a theoretically motivated sociophysicist, the remedy seems to be as 
follows: build a model as simple as possible, control its assumptions, check how the results depend on the parameters, complications 
can be introduced only step by step. This procedure should at least make the model clear to those who deal with real data; it is also in harmony
with warnings, heard from the professional side of sociology \cite{costam}.

On the other hand, the arguments given in our first paragraph suggest that maybe the frames of the game theory are too narrow to describe the 
social entanglement of individual decisions. Between the 'get as much as you can' and the invisible hand of long lasting evolution there is a 
whole spectrum of motivations, not as rational as in economic world, and more rational from an individual or social 
point of view than in the biological evolution. Variable external conditions and individual experience play a role there, but their outcome is 
not fully determined by the expected payoffs. To resolve the puzzle, one should take into account local traditions, norms and social roles 
distribution - a sociophysicist cannot talk much about that. We treat these aspects as a black box. Our input
is the initial distribution of the probabilities of the given strategies; our output is the time dynamics of these probabilities. This dynamics 
is governed by the fundamental or Master equations; this choice seems to be free from the conceptual limitations of the game theory. To maintain the 
continuity with the problem set by Axelrod, we keep the name "the norm game". 

Here we report the results from some models of the norm game which we consider to be as simple as possible, with a minimal
number of parameters. In particular, we reduce chains of reasonings to simple sequences of input and output. 
For example, an agent is expected to cease punishment if the cost of punishment is large. Then we introduce only one parameter to describe how
one act of punishment reduces the probability of punishing; doing this, we omit the information of the punishment cost, its utility etc. 
The method is analytical, and numerical plots are shown for a visualisation. Actually, our aim is to understand the possibilities of our models.
As the basic model here is the mean field approximation, the model can be termed as 'mean-field society'. 

In the following sections, we are going to discuss two models, both of them in two variants. In all cases the results will be the average percentage of punishers and of those who break the norm. Punishing of those who do not punish is not considered here.

\section{Those who break the norm can also punish}

General frames of this model are the same as all models discussed here. The starting points is the Master equation for breaking the norm 
and for punishing for it. What is specific here is that we treat these states as statistically independent. This means 
in particular, that the amount of those who punish among those who obey the norm and those who break it is the same. Intuitively, we could 
expect that those who break the norm do not punish and the opposite. Policemen who drive drunk provide counterexamples; however, this 
case cannot be treated as general one. We start from this model to have a reference point.

Let us denote the probability of breaking the norm as $x$. The time dependence of $x$ is a solution of the Master equation

\begin{equation}
\frac{dz}{dt}=a(1-z)-bzy
\end{equation}
where $y$ is the probability of punishing, $a$ is the rate of increase of $x$ per an agent because of the gain which we get when we break 
the norm and $b$ is the opposite rate because of the inhibiting results of the punishment. We can term these rates as 'temptation' and 'punishment'.
The latter term is proportional to $y$, because a punisher is necessary here. Similarly, the time evolution of $y$ is controlled by 

\begin{equation}
\frac{dy}{dt}=-cyf(z)+ez(1-y)
\end{equation}
where the first term on the r.h.s. is responsible for a reduction of $y$ due to the cost of punishing, and the second - for an increase of $y$
because of growing vengeance - an irritation, that the norm breaking remains non-punished. To fix the timescale, we keep $e=1$ from now on; if we do not,
$c$ should be substituted by $c/e$. It is not clear, if the cost term should  depend on $z$ or not. Does the willingness to punish 
decrease in the absence of those who break the norm? Having no clear answer, we consider two versions of the model: (A1) where $f(z)=z$,
(A2) where $f(z)=1$.

In the case (A1) we get an unique stable solution $z(t)=z^*$, $y(t)=y^*$, where

\begin{equation}
z^*=\big(1+\frac{\phi}{1+c}\big)^{-1}
\end{equation}
\begin{equation}
y^*=(1+c)^{-1}
\end{equation}
where $\phi=b/a$. Then the proportion of the observed behaviour of breaking the norm and punishing is $z^*y^*$, the proportion of breaking the norm 
and not punishing is $z^*(1-y^*)$ and so on. Note that $z^*+y^*\ne 1$, and in general $z+y \ne 1$. 

In the case (A2) we have

\begin{equation}
y^*=\frac{-(1+c)+\sqrt{(1+c)^2+4\phi c}}{2\phi c}
\end{equation}
\begin{equation}
z^*=(1+\phi y^*)^{-1}
\end{equation}
The latter equation is true also in the case (A1). In Fig. 1 we compare the results of two models (A1) and (A2). They do not differ much.

\begin{figure}
\includegraphics[scale=0.55,angle=-90]{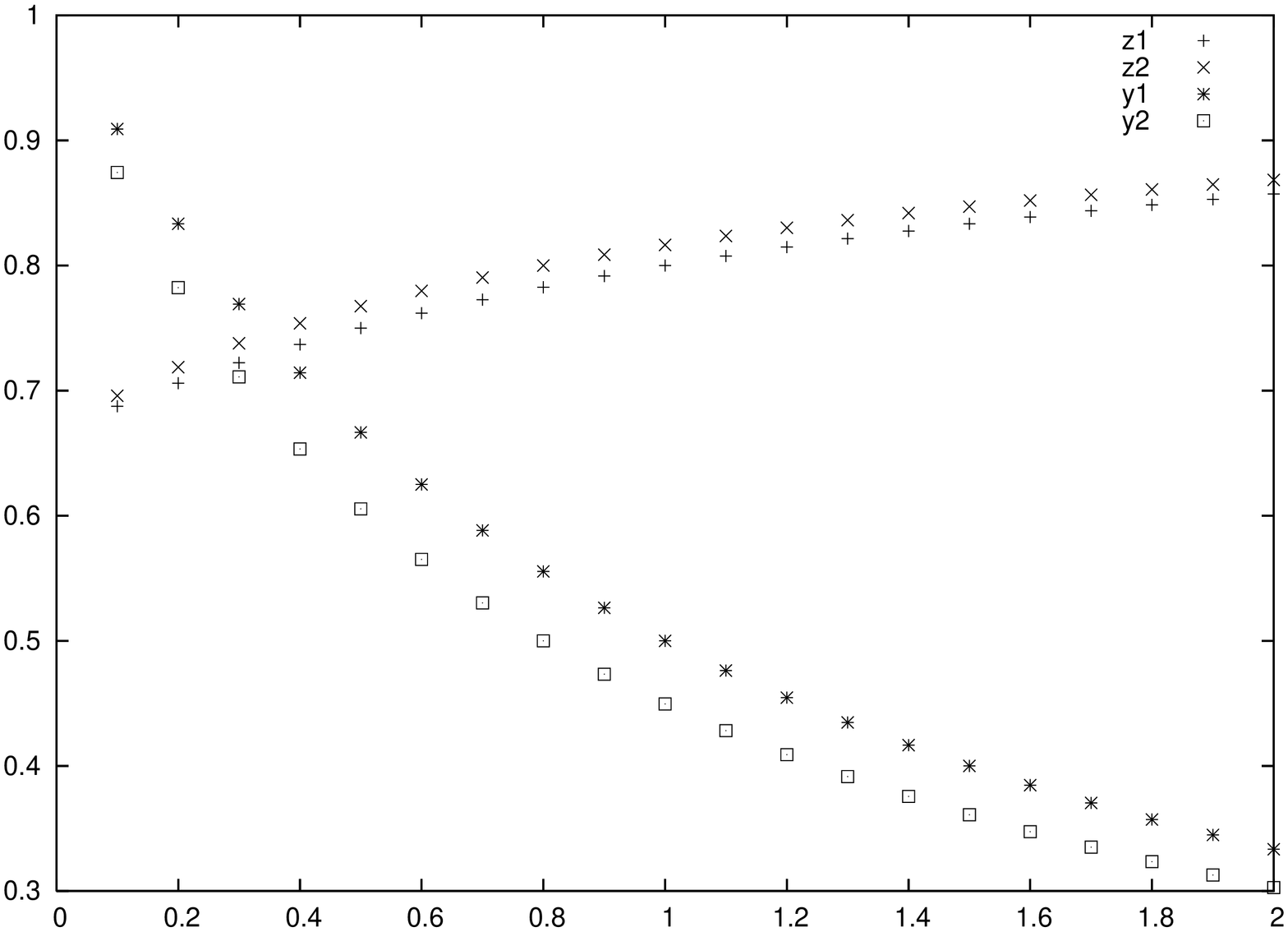}
\includegraphics[scale=0.55,angle=-90]{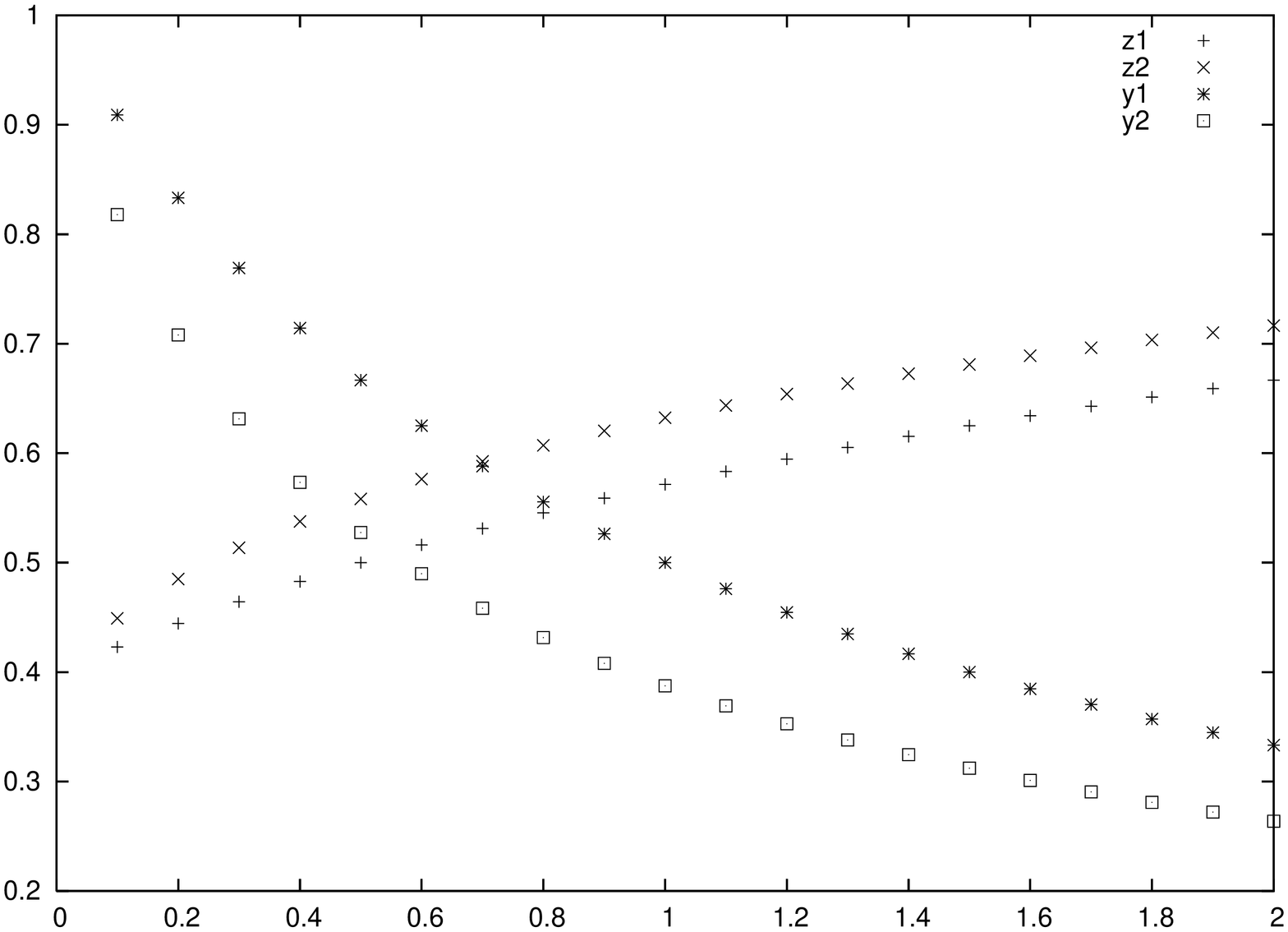}\\
\caption{Comparison of $y^*(c)$ and $z^*(c)$ for the models (A1) and (A2) for {\it a)} $\phi=0.5$, {\it b)} $\phi=1.5$.}
\end{figure}

Let us now consider the case when there are two mutually punishing groups $r$ and $s$. In each of two models (A1) and (A2) there are four equations instead of two, but they are split into two independent pairs. For example, in the model (A1) we have

\begin{equation}
\frac{dz_r}{dt}=a(1-z_r)-bz_ry_s\\
\end{equation}
\begin{equation}
\frac{dy_s}{dt}=-cy_sf(z_r)+z_r(1-y_s)
\end{equation}
plus the same set of equations with the indices $r,s$ interchanged. In this model, those who punish in one group do not contact with those who break the norm in the same group. The differences between the solutions can appear if the values of the parameters are different. For example, a reduction of punishing constant $b$ of one group by the other -- but not the opposite -- can leads to a difference between the solutions $z_r$ and $z_s$. If the parameters $a,b,c$ are the same for both groups, such a difference cannot be obtained.

\section{Only those who obey the norm can punish}

Now we are going to discuss another model, a simplified version of the one presented in Ref. \cite{hau}. Main difference between this and the previous formulations (A1,A2) is that now we do not allow to break the norm and to punish simultaneously.
Then, now there are three possible strategies: {\it i)} to obey the norm and do not punish, {\it ii)} to obey the norm and punish and {\it iii)} to break the norm and not punish. The probabilities of observing these strategies will be denoted as $x$, $y$ and $z$, respectively. Now the normalization condition is $x+y+z=1$. With this condition we can limit the evolution equations to those for $y$ and $z$.

Here again we have two models (B1) and (B2), what is due to the same opportunity. The equations of motion are 

\begin{equation}
\frac{dz}{dt}=ax-bzy 
\end{equation}
\begin{equation}
\frac{dy}{dt}=-cyf(z)+xz
\end{equation}
where as before $f(z)=z$ in (B1) or $f(z)=1$ in (B2). We note that in this model we do not allow to switch directly from $y$ to $z$ or back. This means that all the processes which contribute to changes of $y$ or $z$ do contribute to changes of $x$. 

The solutions show a bifurcation. In the case (B1), a new fixed point appears: $(x^*,y^*,z^*)=(0,0,1)$. This means that everybody breaks the norm and nobody punishes. This solution exists in the whole space $(a,b,c)$, but it is stable if and only if $\phi < c$, i.e. if the punishment constant is small enough. There is also another fixed point $(x^*,y^*,z^*)=(0,1,0)$, but it is never stable. Third solution is

\begin{equation}
z^*=c/\phi
\end{equation}
\begin{equation}
y^*=\frac{1-c/\phi}{1+c}
\end{equation}
and it is stable if and only if $\phi >c$. Moreover, out of this range the probability $y^*$ happens to be negative; then this solution is meaningless.
However, this bifurcation in the two-dimensional space $(z,y)$ is close to what is termed 'transcritical bifurcation' in the case of one variable \cite{glen}.

In the case (B2), the fixed point $(x^*,y^*,z^*)=(0,0,1)$ appears again and it is stable in the same range as in (B1), i.e. when $\phi <c$. The fixed point $(0,1,0)$ does not exist. The other solution is 
\begin{equation}
z^*=\sqrt{c/\phi}
\end{equation}
\begin{equation}
y^*=\frac{1-\sqrt{c/\phi}}{1+\sqrt{c\phi }}
\end{equation}
and it is stable and meaningful again if and only if  $\phi >c$.  As before, we compare the fixed points coordinates in (B1) and (B2) as dependent on the constant $c$; the results are shown in Fig. 2 a. The largest difference is between the curves of $z^*$; for (B1) it is linear, and for (B2) - the square root. Above $c=\phi$, $y^*=0$ and $z^*=1$ in both (B1) and (B2).

\begin{figure}
\includegraphics[scale=0.55,angle=-90]{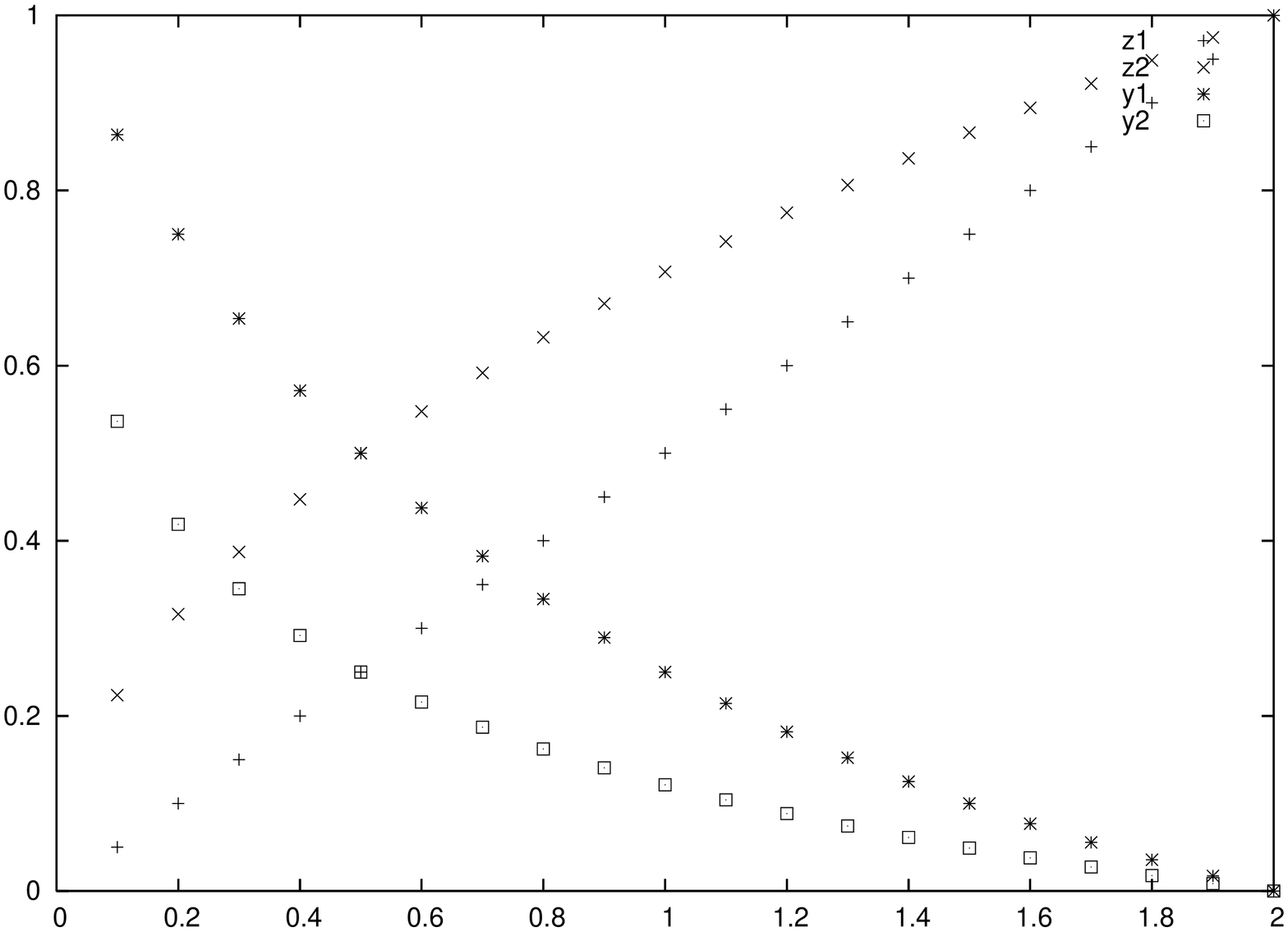}
\includegraphics[scale=0.55,angle=-90]{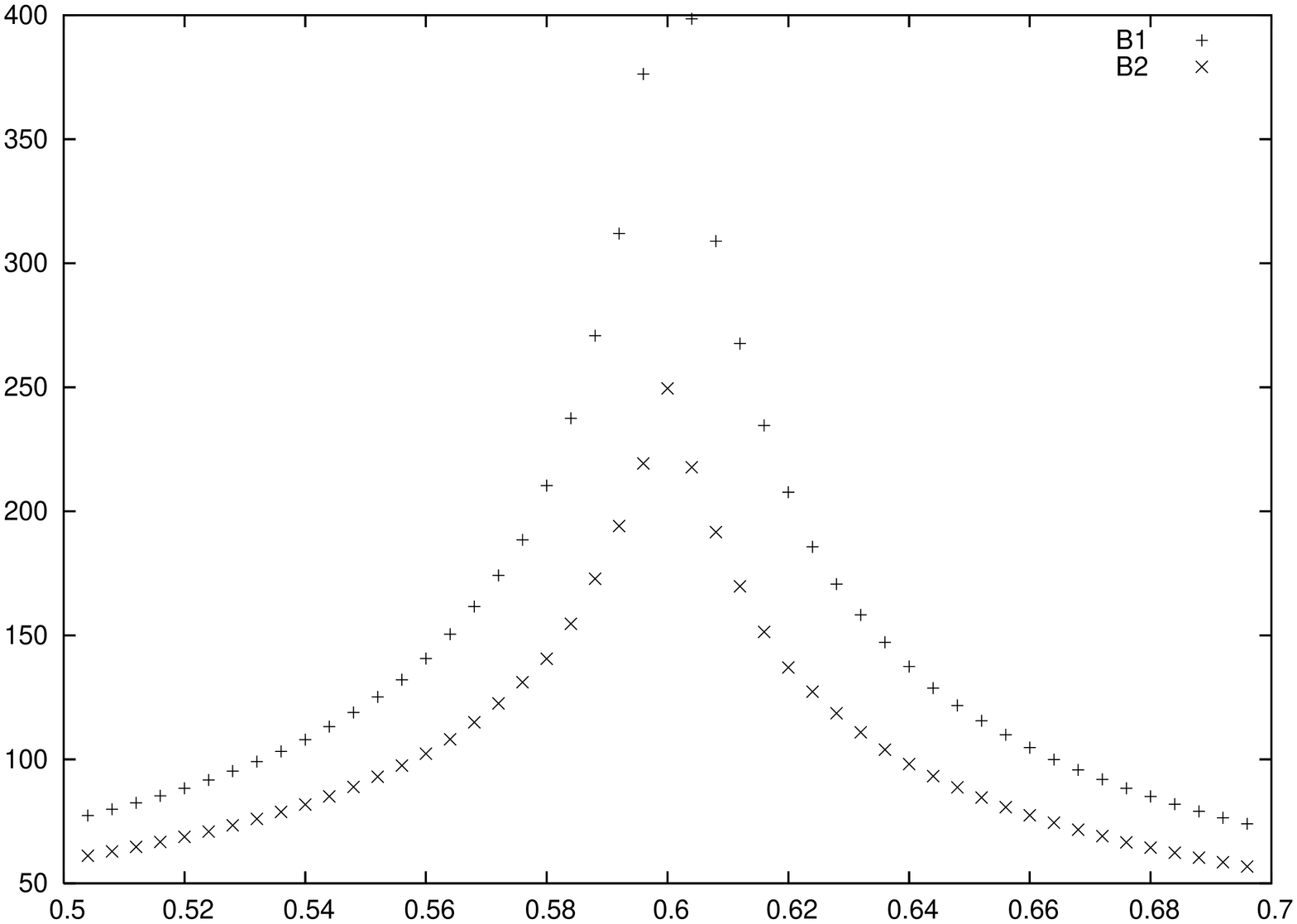}
\caption{{\it a)} Comparison of $y^*(c)$ and $z^*(c)$ for the models (B1) and (B2) for $\phi=2.0$, {\it b)} The time of getting the vicinity of the fixed point, i.e. a sphere with radius $10^{-2}$, as dependent on $\phi$, for the models (B1) and (B2). The parameter $c=0.6$.} 
\end{figure}
 
Once $\phi =c$, solutions given by Eqs. (11,12) (version B1) and (13,14) (version B2) coincide with the branch $(x^*,y^*,z^*)=(0,0,1)$. All these
solutions lose their stability there. This means that one of the eigenvalues $\lambda_1, \lambda_2$ of the Jacobian of partial derivatives of the 
r.h.s of Eqs. (9,10) is zero at the transition point $(x^*,y^*,z^*)=(0,0,1)$. Both in (B1) and (B2), the time of getting equilibrium, defined as $\tau=1/min(-\lambda)$, is infinite at the transition point. Our numerical experiment shows, that near the critical point $\phi=c$ the time of getting the vicinity of the fixed point is remarkably larger for (B1) than for (B2). This effect, shown in Fig. 2 b, cannot be obtained within the standard analysis
of the stability; however, it can be important for applications. In terms of Ref. \cite{costam}, this is a difference which could make a difference.

The case of two mutually punishing groups is described by the equations

\begin{equation}
\frac{dz_i}{dt}=ax_i-bz_iy_{3-i} 
\end{equation}
\begin{equation}
\frac{dy_i}{dt}=-cy_if(z_{3-i})+xz_{3-i}
\end{equation}
where the normalization conditions are $x_i+y_i+z_i=1$ for $i=1,2$. On the contrary to the models (A1,A2) punishers do contact with the rest of their group. 
However, algebraic manipulations show that the fixed points are the same as for the case of one group: $x_1=x_2$, $y_1=y_2$, $z_1=z_2$. The 
stability of the fixed points is to be checked from the eigenvalues of the Jacobian $4\times 4$. In (B1) and at the fixed point $(0,1,0)$ the eigenvalues $\lambda$ fulfil the condition $\lambda (\lambda +a+b)=\pm ac$; one solution is always positive. This means, that the fixed point is never stable. At the fixed point $(0,0,1)$ the secular equation is $(a+\lambda)(1+c+\lambda)-a=\pm b$; all roots are negative if and only if $ac>b$, what coincides with the case of one group. The same is true also for (B2); the only difference is that the fixed point $(0,1,0)$ does not exist. It seems likely, that the fixed
point given by Eqs. (11,12) for (B1) and Eqs. (13,14) for (B2) remains stable also for two groups, if $\phi >c$. This tentative conclusion is confirmed 
by our computer simulation; we have not observed any deviation from the stable fixed point. Then again, a difference in the positions of two groups can appear only if the parameters $a$, $b$ or $c$ are different.

\section{Conclusions}

We have discussed four models. The models (A1,A2) contain the assumption that there is no correlation between obeying or breaking the norm and punishing for breaking it. In the models (B1,B2) we assume that the correlation is strong: nobody who breaks the norm can punish. We expect that any real case is placed between these two extremes. The difference between (A1) and (A2) and between (B1) and (B2) is the same: the relaxation of the probability of punishing does (A1,B1) or does not (A2,B2) depend on the probability that the norm is broken. This difference appears to influence the stationary solution only quantitatively. When we deal with the social reality, these quantitative details can be of minor importance. The difference between the models (A1,A2) and (B1,B2) is more serious; a separate phase appears where the probability of punishment is zero; on the contrary, nobody preserves the norm. This stationary solution appears if the punishment is not effective when compared to the gain for breaking the norm and the cost of punishing. The boundary of this phase in the space of parameters is the condition $b=ac$, as it was discussed in two previous sections. Then the assumption of the correlation between obeying or not the norm and punishing is important for the model results and it would be desirable to check it experimentally before any model is applied to a given norm. Despite the differences, some results are the same in all models. These are: breaking the norm is more frequent when the cost of punishing increases; simultaneously, the punishment itself is less likely to be observed. 

Two patterns of behaviour: to break the norm and to punish, discussed above seem to have some organizing power, attracting collective actions of young
men and adult professionals. Both need a cooperation. On the contrary, those who just obey norms but do not punish often prefer individual activities.
In well-ruled countries there is no overlap between robbers and cops, at least in a social scale. In others, the power means the ability to break
norms and evade being punished. There, the society is inhomogeneous: some obey norms, some do not, and the behaviour is correlated with a place in the social structure. We made a step towards modeling this inhomogeneities, when discussing two mutually punishing groups. Our result is that the positions of the groups are different only if their parameters are different. This fact is reflected in what is observed in political reality, where political parties struggle to make their control from outside (and punishment) less effective, exposing simultaneously events when norms are broken by members of rival parties.


\end{document}